\documentclass[11pt,twoside]{article}
\usepackage{asp2010}
\resetcounters

\begin{document}

\title{Constraints on the Field Star IMF from Resolved Stellar Populations based Star Formation Histories}
\author{Daniel~R.~Weisz$^1$
\affil{$^1$University of Minnesota, Department of Astronomy, 416 Church St SE, Minneapolis, MN 55455}}

\begin{abstract}
Using HST/ACS observations of resolved stellar populations in nearby galaxies, I explore the constraints one can place on the field star IMF from star formation histories (SFHs) derived from synthetic color-magnitude diagram (CMD) fitting.  In particular, I show how reasonable variations in the slope of the IMF, relative to a Salpeter slope, lead to only minor changes in the SFHs.  This shows that CMD-SFH fitting parameter space has a broad minimum with respect to IMF variations and implies that CMD based SFHs can only provide weak explicit constraints on the stellar IMF.   However, observations of resolved stellar populations in dwarf galaxies can be used to tandem with other methods to search for variations in the upper IMF.
\end{abstract}

\section{Introduction}

The properties of massive stars are critical to much of what we understand about the formation and evolution of galaxies.  These luminous stars often account for a significant fraction of the total luminosity of galaxies, and are critical to determining physical parameters such as stellar mass or star formation rate (SFR).  At the core of interpreting the luminosities of massive star is the upper mass IMF, which assigns a distribution of masses to a stellar population.   

The gold standard IMF has long been the universal Salpeter IMF \citep{sal55}. However, recent studies in nearby galaxies have found a lower than expected H$\alpha$ fluxes in low mass galaxies \citep[e.g.,][]{sul04, igl04,hov08,bos09, lee09,meu09}, leading some to suggest that the IMF may vary with respect to environment \citep[e.g.,][]{hov08,lee09,meu09, pfl09}.  Others contend that highly variable SFHs could equally well explain the observed trend \citep[e.g.,][]{sul04, bos09}.  

With Hubble Space Telescope optical imaging of nearly $\sim$ 70 galaxies in the Local Volume, the ACS Nearby Galaxy Survey Treasury program \citep[ANGST;][]{dal09} provides high quality photometric measurements of resolved stellar populations.  It is often assumed that such a dataset can shed light into the the SFH-IMF debate, either via a census of massive stars or by searching for the most likely SFHs over a range of input IMF slopes.  In light of this, I present empirical analysis of ANGST data and discuss the potential uses and pitfalls of resolved stellar populations data for understanding potential IMF variations in dwarf galaxies.

\section{Constraints on the IMF from Color-Magnitude Diagrams and Star Formation Histories}

Optical imaging of resolved stellar populations in nearby galaxies allows us to construct color-magnitude diagrams (CMDs), which contain a record of all star formation that has ever occurred in that galaxy (Figure \ref{fig:ddo53imf}).    A number of sophisticated algorithms have been developed to extract the SFHs from observed CMDs by numerical comparison with simulated CMDs (e.g., \citealt{gal05} and references therein) .

For the ANGST program, I have selected one such code described in \citet{dol02}.  Using this method, a user specifies an assumed IMF and binary fraction, and allowable ranges in age, metallicity, distance, and extinction.  Photometric errors and completeness are characterized by artificial star tests.  From these inputs, many synthetic CMDs are generated to span the desired age and metallicity range.  For this work, I have used synthetic CMDs sampling stars with age and metallicity spreads of 0.1 dex.  These individual synthetic CMDs are then linearly combined along with a model foreground CMD to produce a composite synthetic CMD.  The linear weights on the individual CMDs are adjusted to obtain the best fit as measured by a Poisson maximum likelihood statistic; the weights corresponding to the best fit are the most probable SFH.  This process can be repeated at a variety of distance and extinction values to solve for these parameters as well.

I emphasize that the IMF in the process of measuring SFHs is a changeable input parameter, and can thus be varied to explore the effects of different choices of the IMF on the resultant SFHs.

\subsection{Empirically Exploring the SFH-IMF Degeneracy}

While it is well known SFHs and the IMF are degenerate \citep[e.g.,][]{mil79, elm06}, the degree to which the IMF and SFHs are degenerate has not been extensively explored in the context of observational data.   As a demonstration of the empirically derived degeneracy, I selected a representative galaxy from the ANGST sample, M81~Group dwarf irregular galaxy DDO~53 (UGC~4459).  Using the ANGST photometry and false stars as input, I assumed a TRGB-based distance modulus of 27.29 \citep{dal09} and a foreground extinction value of $A_V$=0.12 \citep{sch98}.  For the first SFH solution I chose a single-sloped Salpeter IMF ($x$=1.35).  I then systematically varied the assumed IMF for each subsequent solution.  For each best fit CMD, I calculated the reduced $\chi ^{2}$ value, a measure of the goodness-of-fit.

\begin{figure}[h]
\plottwo{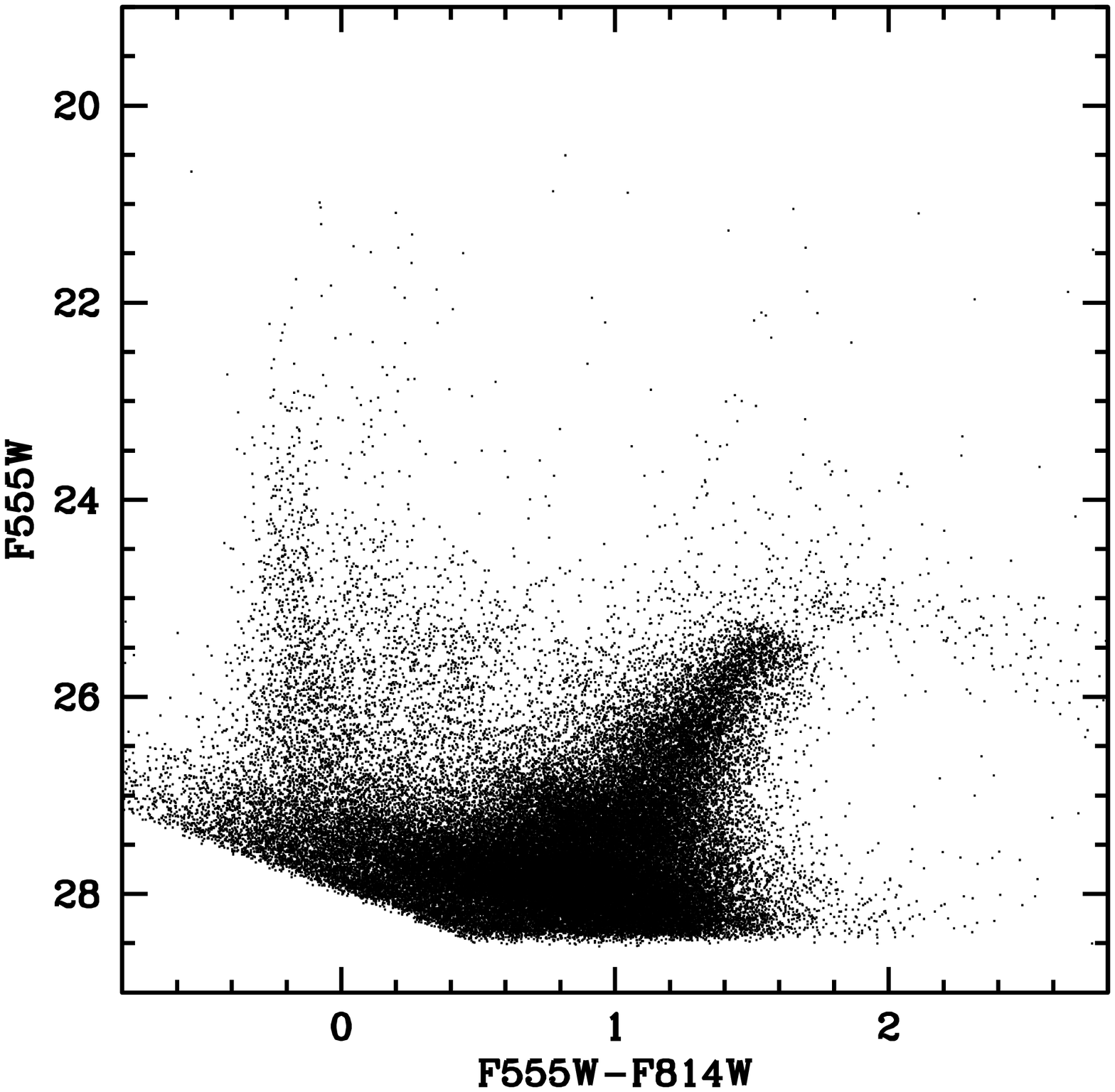}{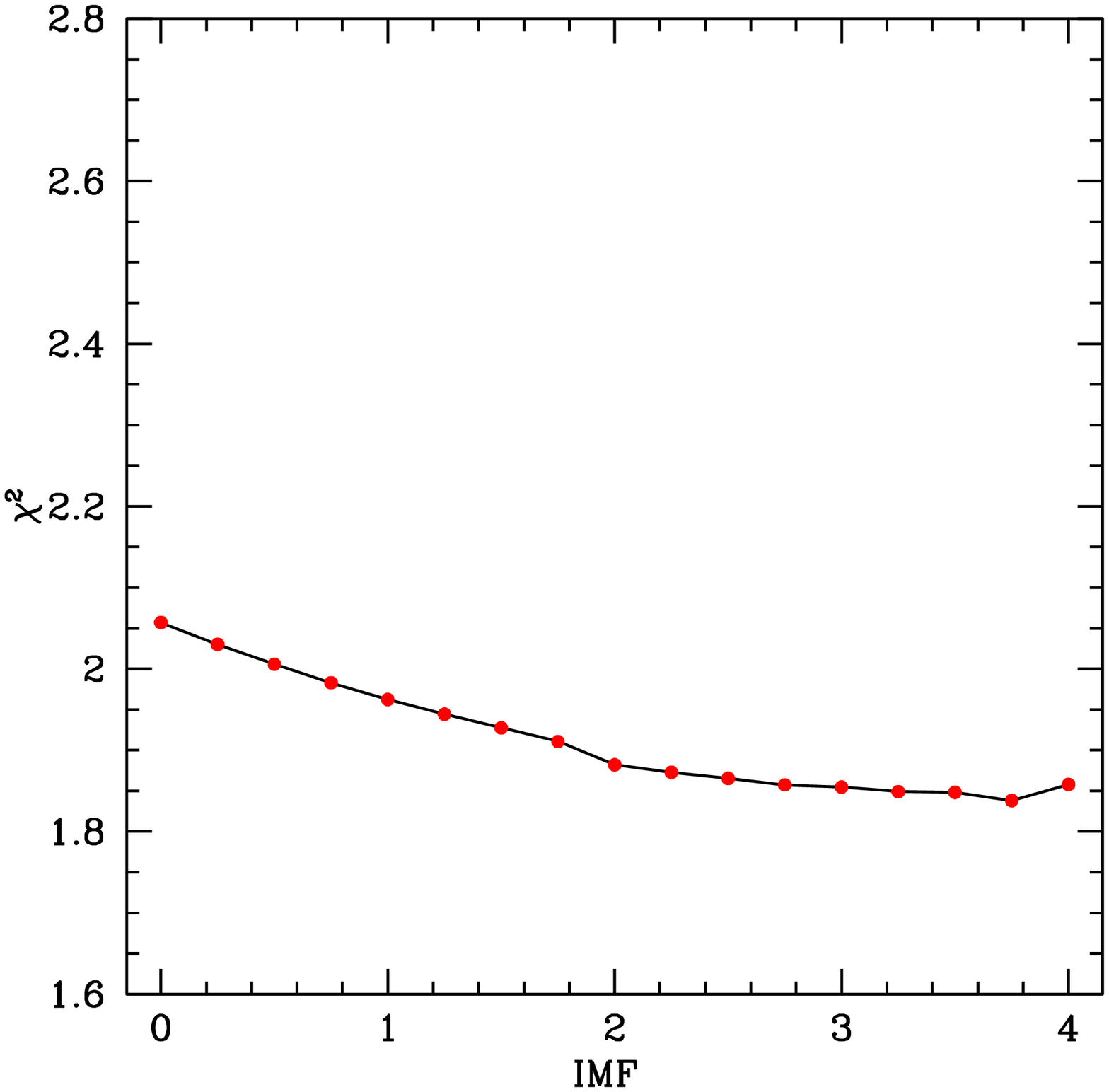}
\caption{Left Panel--The observed CMD of M81~Group dwarf irregular galaxy DDO~53 from the ANGST program.  Right panel--The reduced $\chi^2$ parameter from the most likely fit SFH versus the choice of input IMF.  In this plot, a Salpeter IMF has a value of 1.35.  Note the small amplitude of variations in $\chi^2$ for a wide range of selected IMF slopes, which empirically demonstrates the SFH-IMF degeneracy.
\label{fig:ddo53imf}
}
\end{figure}

Figure \ref{fig:ddo53imf} illustrates the broad $\chi^2$ minimum over a wide range of selected IMF slopes. This finding empirically reinforces the expected nature of the SFH-IMF degeneracy, namely that a number of SFHs are nearly equally well fit for a range of assumed IMFs.   It is important to note that variations in the $\chi ^2$ values are not necessarily attributable to just the change in IMF.  Indeed, variations in the metallicity and isochronal degeneracies, e.g., different age MS stars can have the same optical colors and magnitudes, also can contribute to these variations.  

\subsection{The Effects of Varying the IMF on Simulated CMDs}

To illustrate the effects of different choices of IMFs on a stellar population, I use the diagnostic power of simulated CMDs.  Assuming a Salpeter IMF, I solved for the SFH and chemical evolution history of DDO~53.  I then used the derived SFH and chemical evolution law as inputs into creating a simulated CMD.  That is, in this case, I assume prior knowledge of the SFH, and hold it fixed, as I vary the IMF.

\begin{figure}[h!]
\plotone{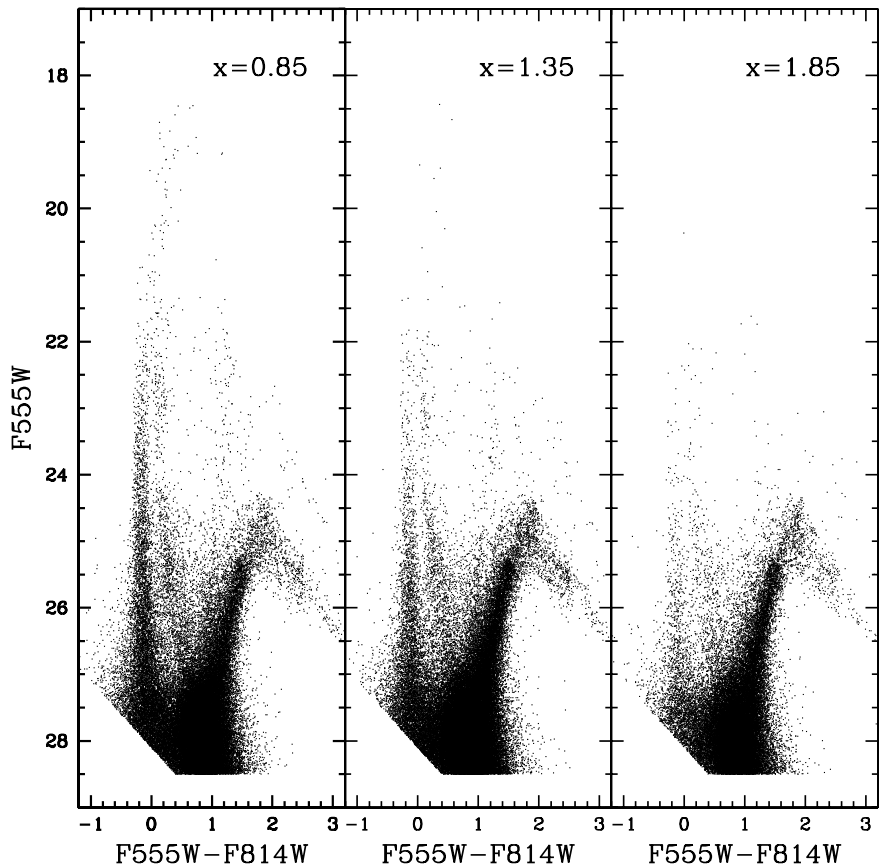}
\caption{Simulated CMDs constructed from a fixed SFH, i.e., the one measured from the observed CMD using a Salpeter IMF, but with different choices for input IMF parameters.  Note that varying the IMF from flatter to steeper causes a drastic reduction in the number of luminous young stars.  This effect could also be explained by different SFHs, demonstrating the essence of the degeneracy.
\label{fig:ddo53sim}
}
\end{figure}

Figure \ref{fig:ddo53sim} shows selected simulated CMDs created with different choices for the IMF.  The left panel shows a CMD for a flatter-than-Salpeter IMF ($x$=0.85) , the central panel a Salpeter IMF ($x$=1.35) , and the right panel a steeper-than-Salpeter IMF($x$=1.85).  At first glance, is appears that these CMD could simply represent different galaxies, i.e., different SFHs.  The flatter IMF has a larger number of luminous young stars, which could also be achieved by assuming a Salpeter IMF with an increased recent SFR.  The choice of a steeper IMF results in less young, luminous stars, and the resultant simulated CMD could be consistent with a `post-burst' SFH.  An even steeper assumed IMF would transform this into something akin to a purely old stellar population.  Our ability to tell the difference between the effects of SFH and IMF variations are extremely limited for mixed age stellar populations.  One alternative is to examine stellar clusters, where the SFH can assumed to be single-aged, thus breaking the degeneracy.

\section{ANGST Clusters and the IMF}

In principle, the coeval nature of stellar clusters provide an ideal place to test for variations in the IMF.  Photometric tests of this nature have been performed in the Local Group \citep[e.g.,][]{hil94}, with mixed success, typically due to a small number of sample clusters.   While the ANGST sample likely contains hundreds or more young stellar clusters that span a range in environments, even the high spatial resolution afforded by HST cannot resolve the individual stars in all clusters.  For example, at the distance of DDO~53, a single ACS pixel is $\sim$ 1 pc in physical size.  Although cluster sizes can range from sub-pc to hundreds of pc in size, the central regions of young clusters typically appear as blended even in HST photometry.  Further, photometric quality cuts (e.g., crowding, sharpness) can influence the selection of stars, often at the expense of luminous objects \citep[e.g.,][]{gog09}, resulting in an undercounting of luminous stars in most nearby galaxies.  

\section{Potential Uses of Resolved Stars for Constraining the Upper IMF}

Explicitly placing constraints on the slope of the upper IMF is not a forte of resolved stellar populations in galaxies outside the Local Group. However, high fidelity observations of resolved stellar populations can be used in tandem with other methods.  For example, the one of the prominent issues in H$\alpha$-FUV flux ratio studies is the percentage of ionizing photons that do not result in H$\alpha$ emission, but instead escape the galaxy without scattering.  This issue of `leakage' could be significant enough to explain lower than expected H$\alpha$ in dwarf galaxies \citep[e.g.,][]{bos09, lee09, meu09}.  Examining the resolved stellar populations of dwarf galaxies could reveal the location of field O stars, which either may have associated H$\alpha$ emission below the current detection limits, or not have association H$\alpha$ at all.  Such a catalog of O stars could be used as a guide for followup deep H$\alpha$ imaging.

The varying IMF interpretation of H$\alpha$-FUV flux ratio studies relies upon the assumption that modeled SFHs cannot account for the observed trend.  However, a number of studies did not include models that resemble the observed SFHs of dwarf galaxies, which are typically highly variable over short time scales \citep[$\lesssim$ 100 Myr; e.g.,][]{dom97,dol05,wei08,mcq09}.  In contrast, the study by \citet{meu09} included a single large amplitude Gaussian burst with a FWHM $\sim$ 1 Gyr.  Similarly, the work by \citet{hov08} considered a large number of models, however the  typical dwarf galaxy model consisted bursts of $\sim$ 200 Myr in duration with mean spacings of $\sim$ 1 Gyr, which is not consistent with measured SFHs in nearby dwarf galaxies.  The inclusion of realistic modeled SFHs is essential to accurately interpreting the lower than expected H$\alpha$-FUV flux ratios in nearby galaxies.

\section{Summary}

Although HST observations of resolved stellar populations seem in nearby dwarf galaxies to be an ideal dataset for exploring variations in the upper IMF, there are several challenges which make an explicit study of the IMF untenable.  The first difficulty arises due to the SFH-IMF degeneracy.  From empirical analysis of ANGST dwarf irregular galaxy DDO~53, I demonstrated that a deficit of massive stars on the CMD could be due to either a steeper than Salpeter IMF or a lack of recent SFH -- the two scenarios are indistinguishable.  Coeval stellar clusters provide the opportunity to explore IMF variations as the SFH is known \emph{a prioi}.  However, photometric crowding and blending of individual stars for galaxies located outside the LG result in an under counting of luminous young stars in most cases.  

Resolved stellar populations provide two clear indirect ways of helping constrain the upper IMF.  From the CMDs, one can identify the location of potential field O stars, which could contribute to undetected low surface brightness H$\alpha$ or simply have no gaseous component and thus contribute to ionizing photon `leakage'.  Further, the SFHs derived from resolved stellar populations can be used as templates for models SFHs considered in H$\alpha$-FUV flux ratio studies.  These studies are heavily dependent upon the assumed SFHs, and the inclusions of realistic SFHs of dwarf galaxies is essential to a robust interpretation of the results.

\acknowledgements DRW would like to thank Evan Skillman, Nate Bastian, Daniela Calzetti, and Cliff Johnson for their insightful comments and discussion.  DRW is also grateful for support from the University of Minnesota Doctoral Dissertation Fellowship and Penrose Fellowship. This work is based on observations made with the NASA/ESA Hubble Space Telescope, obtained from the data archive at the Space Telescope Science Institute.  Support for this work was provided by NASA through grants GO-10915, DD-11307, and GO-11986 from the Space Telescope Science Institute, which is operated by AURA, Inc., under NASA contract NAS5-26555. This research has made use of the NASA/IPAC Extragalactic Database (NED), which is operated by the Jet Propulsion Laboratory, California Institute of Technology, under contract with the National Aeronautics and Space Administration.  This research has made extensive use of NASA's Astrophysics Data System Bibliographic Services.  

\bibliography{weisz_d}

\end{document}